\documentclass[preprint,aps]{revtex4}
\newcommand{\be}{\begin{equation}}
\newcommand{\ee}{\end{equation}}
\newcommand{\bea}{\begin{eqnarray}}
\newcommand{\eea}{\end{eqnarray}}
\newcommand{\ba}{\begin{array}}
\newcommand{\ea}{\end{array}}

\begin{document}
\title{Spin and Flavor Strange Quark Content of the Nucleon}
\author{Harleen Dahiya$^a$  and Manmohan Gupta$^b$}
\affiliation{$^a$ Department of Physics, Dr. B.R. Ambedkar
National Institute of Technology, Jalandhar, 144001, India\\ $^b$
Department of Physics, Centre of Advanced Study in Physics, Panjab
University, Chandigarh 160014, India}
\date{\today}

\begin{abstract}
Several spin and flavor dependent parameters characterizing the
strangeness content of the nucleon have been calculated in the
chiral constituent quark model with configuration mixing
($\chi$CQM$_{{\rm config}}$) which is known to provide a
satisfactory explanation of the ``proton spin crisis'' and related
issues. In particular, we have calculated the strange spin
polarization $\Delta s$, the strangeness contribution to the weak
axial vector couplings $\Delta_8$ etc., strangeness contribution
to the magnetic moments $\mu(p)^s$ etc., the strange quark flavor
fraction $f_s$, the strangeness dependent quark flavor ratios
$\frac{2 \bar s}{u+d}$ and $\frac{2 \bar s}{\bar u+ \bar d}$ etc..
Our results are consistent with the recent experimental
observations.
\end{abstract}

\maketitle

The recent measurements by  several groups SAMPLE  at MIT-Bates
\cite{sample}, G0  at JLab \cite{g0}, A4  at MAMI \cite{a4} and by
HAPPEX  at JLab \cite{happex} regarding the contribution of
strangeness to the electromagnetic form factors of the nucleon
have triggered a great deal of interest in finding the strangeness
magnetic moment of the proton ($\mu(p)^s$). The SAMPLE experiment
has observed $\mu(p)^s$ to be $0.37 \pm 0.26 \pm 0.20$
\cite{sample} whereas G0 \cite{g0}, A4 \cite{a4} and HAPPEX
\cite{happex} have observed the combination of electric and
magnetic form factors. A global fit  has also been carried out by
including the data from SAMPLE, G0, A4 and HAPPEX and it gives
$\mu(p)^s=0.12 \pm 0.55 \pm 0.07$ \cite{global}. The first lattice
calculations by Dong {\it et al.} give $-0.36 \pm 0.20$
\cite{lattice} whereas a recent phenomenology study with lattice
inputs \cite{mup-aw} also predicts a very small value for the
strange magnetic moment, $\mu(p)^s=-0.046 \pm 0.19 \mu_N$. In the
naive constituent quark model (CQM) \cite{dgg,Isgur,yaouanc},
$\mu(p)^s$ is predicted to be zero. The broader question of
contribution of strangeness in the nucleon has also been discussed
by several authors recently \cite{ellis-strange}. It is widely
recognized that a knowledge about the strangeness content of the
nucleon would undoubtedly provide vital clues to the
non-perturbative aspects of QCD.

The existence of strangeness in the nucleon has been indicated in
the context of low energy experiments \cite{nutev,sigma} whereas
it has been observed in the deep inelastic scattering (DIS)
experiments \cite{emc,smc,adams,hermes}. In the context of DIS,
the strange spin polarization of the nucleon \cite{ellis-kar}
looks to be well established through the measurements of polarized
structure functions of the nucleon \cite{smc,adams,hermes}. Apart
from the observations of DIS data regarding strangeness dependent
spin polarization functions, several interesting facts have also
been revealed regarding the quark flavor distribution functions in
the DIS experiments. In particular, the NuSea Collaboration
\cite{e866} have given the results for the  integrals of $\bar
u-\bar d$ and $\bar u/\bar d$ asymmetries indicating that the
flavor structure of the nucleon is not limited to $u$ and $d$
quarks only.  Also the CCFR Collaboration and more recently the
NuTeV collaboration \cite{nutev} have given fairly good deal of
information regarding the integrals of strangeness dependent quark
ratios in the nucleon given as $\frac{2 \bar s}{u+d}
=0.099^{+0.009}_{-0.006}$ and $\frac{2\bar s}{\bar u+\bar d}
=0.477^{+0.063}_{-0.053}.$ In the context of low energy
experiments, the large pion-nucleon sigma term value \cite{sigma}
indicating non zero strange quark flavor fraction $f_s$ is also
indicative of the presence of strange quarks in the nucleon
although there is no consensus regarding the various mechanisms
which can contribute to  $f_s$ \cite{gasser}. Therefore, the
indications of the strange quark degree of freedom in DIS as well
as low energy experiments provide a strong motivation to examine
the strangeness contribution to the nucleon thereby giving vital
clues to the non-perturbative effects of QCD.

One may think that the strangeness content of the nucleon perhaps
can be obtained through the generation of ``quark sea''
perturbatively from the quark-pair production by gluons. However,
this kind of ``sea'' is symmetric w.r.t. $\bar u$ and $\bar d$
\cite{cheng}, negated by the observed value of $\bar u-\bar d$
asymmetry \cite{e866}. Therefore, one has to consider the ``quark
sea'' produced by the non-perturbative mechanism. One such model
which can yield an adequate description of the ``quark sea''
generation through the chiral fluctuations is the chiral
constituent quark model ($\chi$CQM) \cite{manohar} which is not
only successful in giving a satisfactory explanation of ``proton
spin crisis'' \cite{eichten} but is also able to account for the
violation of Gottfried Sum Rule \cite{cheng,eichten,GSR}, baryon
magnetic moments and hyperon $\beta-$decay parameters
\cite{cheng,johan}. Recently, it has been shown that configuration
mixing generated by spin-spin forces improves the predictions of
$\chi$CQM regarding the quark distribution functions and spin
polarization functions \cite{hd}. Further, the chiral constituent
quark model with configuration mixing ($\chi$CQM$_{{\rm config}}$)
when coupled with the quark sea polarization and orbital angular
momentum through the Cheng-Li mechansim \cite{cheng} is able to
give an excellent fit \cite{hdorbit} to the octet magnetic
moments. It, therefore, becomes desirable to carry out a detailed
analysis in the $\chi$CQM$_{{\rm config}}$ of the strangeness
dependent spin polarization functions as well as the quark
distribution functions, particularly in the light of some recent
observations \cite{sample,g0,a4,happex,nutev,sigma,adams,bass}.

The purpose of the present communication is to carry out detailed
calculations of the parameters  characterizing the strangeness of
the nucleon within the $\chi$CQM$_{{\rm config}}$. In particular,
we would like to calculate the strange spin polarization $\Delta
s$, strangeness contribution to the weak axial vector couplings
$\Delta_3$, $\Delta_8$ and $\Delta_0$, strangeness contribution to
the magnetic moments $\mu(p)^s$ and $\mu(n)^s$,  the strange quark
flavor fraction $f_s$,  the strangeness dependent quark flavor
ratios $\frac{2 \bar s}{u+d}$ and $\frac{2 \bar s}{\bar u+ \bar
d}$. For the sake of completeness, we would also like to calculate
the strangeness contribution to the magnetic moments of decuplet
baryons $\mu(\Delta^{++})^s$, $\mu(\Delta^{+})^s$,
$\mu(\Delta^{o})^s$ and $\mu(\Delta^{-})^s$ which have not been
observed experimentally.

To make the mss. more readable as well as for ready reference, we
mention the essentials of $\chi$CQM$_{{\rm config}}$, for details
we refer the reader to \cite{cheng,johan,hd}. The basic process in
the $\chi$CQM formalism is the emission of a Goldstone boson (GB)
by a constituent quark which further splits into a $q \bar q$
pair, for example,

\be
  q_{\pm} \rightarrow {\rm GB}^{0}
  + q^{'}_{\mp} \rightarrow  (q \bar q^{'})
  +q_{\mp}^{'}\,,                              \label{basic}
\ee where $q \bar q^{'} +q^{'}$ constitute the ``quark sea''
\cite{cheng} and the $\pm$ signs refer to the quark helicities.
The effective Lagrangian describing interaction between quarks and
a nonet of GBs, consisting of octet and a singlet, can be
expressed as
\be
{\cal L}= g_8 {\bf \bar q}\left(\Phi+\zeta\frac{\eta'}{\sqrt 3}I
\right) {\bf q}=g_8 {\bf \bar q}\left(\Phi'\right) {\bf q}\,, \ee
where  $\zeta=g_1/g_8$, $g_1$ and $g_8$ are the coupling constants
for the singlet and octet GBs, respectively, $I$ is the $3\times
3$ identity matrix. The GB field which includes the octet and the
singlet GBs is written as \bea
 \Phi' = \left( \ba{ccc} \frac{\pi^0}{\sqrt 2}
+\beta\frac{\eta}{\sqrt 6}+\zeta\frac{\eta^{'}}{\sqrt 3} & \pi^+
  & \alpha K^+   \\
\pi^- & -\frac{\pi^0}{\sqrt 2} +\beta \frac{\eta}{\sqrt 6}
+\zeta\frac{\eta^{'}}{\sqrt 3}  &  \alpha K^0  \\
 \alpha K^-  &  \alpha \bar{K}^0  &  -\beta \frac{2\eta}{\sqrt 6}
 +\zeta\frac{\eta^{'}}{\sqrt 3} \ea \right) {\rm and} ~~~~q =\left( \ba{c} u \\ d \\ s \ea
\right)\,. \eea

SU(3) symmetry breaking is introduced by considering $M_s >
M_{u,d}$ as well as by considering the masses of GBs to be
nondegenerate
 $(M_{K,\eta} > M_{\pi}$ and $M_{\eta^{'}} > M_{K,\eta})$
\cite{cheng,johan}. The parameter $a(=|g_8|^2$) denotes the
probability of chiral fluctuation  $u(d) \rightarrow d(u) +
\pi^{+(-)}$, $\alpha^2 a$, $\beta^2 a$ and $\zeta^2 a$
 respectively denote the probabilities of fluctuations
$u(d) \rightarrow s + K^{-(0)}$,  $u(d,s) \rightarrow u(d,s) +
\eta$ and $u(d,s) \rightarrow u(d,s) + \eta^{'}$.  Further, to
make the transition from $\chi$CQM to $\chi$CQM$_{{\rm config}}$,
the nucleon wavefunction gets modified because of the
configuration mixing generated by the chromodynamic spin-spin
forces \cite{{dgg},{Isgur},{yaouanc},hd} as follows,
\begin{equation}
|B\rangle = \cos \phi |56,0^+\rangle_{N=0} + \sin
\phi|70,0^+\rangle_{N=2}\,, \label{mixed}
\end{equation}
where $\phi$ represents the $|56\rangle-|70\rangle$ mixing. For
details of the  spin, isospin and spatial parts of the
wavefunction,  we  refer the reader to {\cite{{yaoubook}}.

Before proceeding further, we briefly discuss the strangeness
dependent spin polarization functions, quark distribution
functions and the related quantities of the nucleon. To begin
with, we consider the spin structure of a nucleon defined as
\cite{{cheng}}
\be
\hat B \equiv \langle B|N|B\rangle, \ee where $|B\rangle$ is the
nucleon wavefunction defined in Eq. (\ref{mixed}) and $N$ is the
number operator given by
\be
 N=n_{u_{+}}u_{+} + n_{u_{-}}u_{-} +
n_{d_{+}}d_{+} + n_{d_{-}}d_{-} + n_{s_{+}}s_{+} +
n_{s_{-}}s_{-}\,, \ee  $n_{q_{\pm}}$ being the number of $q_{\pm}$
quarks. Following Ref. \cite{hd}, the contribution to the proton
spin by different quark flavors in $\chi$CQM$_{{\rm config}}$ can
be given by the spin polarizations $\Delta q=q_+-q_-$ expressed as
\be
   \Delta u =\cos^2 \phi \left[\frac{4}{3}-\frac{a}{3}
   (7+4 \alpha^2+ \frac{4}{3} \beta^2
   + \frac{8}{3} \zeta^2)\right]+ \sin^2 \phi \left[\frac{2}{3}-\frac{a}{3} (5+2 \alpha^2+
  \frac{2}{3} \beta^2 + \frac{4}{3} \zeta^2)\right], \label {du}\ee
\be
  \Delta d =\cos^2 \phi \left[-\frac{1}{3}-\frac{a}{3} (2-\alpha^2-
  \frac{1}{3}\beta^2- \frac{2}{3} \zeta^2)\right]
+ \sin^2 \phi \left[\frac{1}{3}-\frac{a}{3} (4+\alpha^2+
  \frac{1}{3} \beta^2 + \frac{2}{3} \zeta^2)\right], \label{dd}
  \ee
\be   \Delta s =\cos^2 \phi \left[ -a \alpha^2 \right]+\sin^2 \phi
\left[ -a \alpha^2 \right]\,. \ee A closer look at the above
equations reveals that $\phi$ represents the configuration mixing
angle, the constant factors represent the CQM results and the
factors which are multiple of $a$ represent the contribution from
the ``quark sea''. It is important to mention here that the
presence of $s \bar s$ in the ``quark sea'' (Eq. (\ref{basic}))
involves the terms $\alpha^2 a$, $\beta^2 a$ and $\zeta^2 a$
respectively denoting the fluctuations $u(d) \rightarrow s +
K^{+(0)}$, $u(d,s) \rightarrow u(d,s) + \eta$ and $u(d,s)
\rightarrow u(d,s) + \eta^{'}$.  It is clear from the above
fluctuations that the strange quarks come from the fluctuations of
the $u$ and $d$ quarks as well and therefore, apart from
contributing to the strange spin polarization, the strange quarks
also contribute to the spin polarizations of $u$ and $d$ quarks.
It should also be noted that the $\bar d-\bar u$ asymmetry in this
case can be easily understood in terms of the above fluctuations
where the valence $u$ quarks are likely to produce more $\bar d$
than the valence $d$ quarks producing $\bar u$ in the proton. The
spin polarization functions are also related to the axial vector
couplings measured in the baryon weak decays \cite{ellis-kar}, for
example, the non-singlet combinations of the quark spin
polarizations ($\Delta_3$ and $\Delta_8$) can be expressed as \bea
\Delta_3&=& \Delta u-\Delta d\,=F+D\,,
\\ \Delta_8&=& \Delta u+\Delta d-2 \Delta s=3F-D\,, \eea
where $F$ and $D$ are the usual SU(3) parameters characterizing
the weak matrix elements. The flavor singlet combination on the
other hand can be related to the total spin carried by the quarks
as \bea \Delta_0&=&\frac{1}{2} \Delta \Sigma= \frac{1}{2}(\Delta
u+\Delta d+\Delta s)\,. \eea

After discussing the spin polarization functions, we would like to
discuss the formalism for the strangeness contribution to the
magnetic moments. It would be important to mention here that the
predictions for all the octet and decuplet baryon magnetic moments
including the transition magnetic moments  have been discussed in
detail in Ref. \cite{hdorbit} and have been found to be in good
agreement with the recent measurements as well as other
theoretical estimates. The magnetic moment of a given baryon in
the $\chi$CQM can be expressed as
\be
\mu(B)_{{\rm total}} = \mu(B)_{{\rm val}} + \mu(B)_{{\rm sea}}\,,
\ee where $\mu(B)_{{\rm val}}$ represents the contribution of the
valence quarks and $\mu(B)_{{\rm sea}}$ corresponding to the
``quark sea'' (Eq. (\ref{basic})). Further, $\mu(B)_{{\rm sea}}$
can be written as
\be
\mu(B)_{{\rm sea}} = \mu(B)_{{\rm spin}} + \mu(B)_{{\rm orbit}}\,,
\ee where the first term is the magnetic moment contribution of
the $q^{'}$ in Eq. (\ref{basic}) coming from the spin polarization
and the second term is due to the rotational motion of the two
bodies, $q^{'}$ and GB, corresponding to the fluctuation given in
Eq. (\ref{basic}) and referred to as the orbital angular momentum
by Cheng and Li \cite{cheng}.

To find the strangeness contribution to the magnetic moment of the
proton $\mu(p)^s$ we should note that there are no `strange'
valence quarks, therefore $\mu(p)^s$ receives contributions only
from the ``quark sea" and is expressed as
\be
\mu(p)^s = \mu(p)^s_{{\rm spin}} + \mu(p)^s_{{\rm orbit}}\,.
\label{totalmag} \ee Following Ref. \cite{hdorbit},
$\mu(p)^s_{{\rm spin}}$ in the chiral constituent quark model with
configuration mixing can be expressed as
\be
\mu(p)^s_{{\rm spin}}=\sum_{q=u,d,s}\Delta q(p)^s_{{\rm sea}}\mu_q
\,, \label{magsea} \ee
where $\mu_q= \frac{e_q}{2 M_q}$ ($q=u,d,s$) is the quark magnetic
moment, $e_q$ and $M_q$ are the electric charge and the mass
respectively for the quark $q$. The contribution of strangeness to
the spin polarization functions is given as
\be
   \Delta u(p)^s_{{\rm sea}} =\cos^2 \phi \left[-\frac{a}{3}
   (4 \alpha^2+ \frac{4}{3} \beta^2
   + \frac{8}{3} \zeta^2)\right]+ \sin^2 \phi \left[-\frac{a}{3} (2 \alpha^2+
  \frac{2}{3} \beta^2 + \frac{4}{3} \zeta^2)\right], \label{dusea}\ee
\be
  \Delta d(p)^s_{{\rm sea}}=\cos^2 \phi \left[-\frac{a}{3} (-\alpha^2-
  \frac{1}{3}\beta^2- \frac{2}{3} \zeta^2)\right]
+ \sin^2 \phi \left[-\frac{a}{3} (\alpha^2+
  \frac{1}{3} \beta^2 + \frac{2}{3} \zeta^2)\right],
  \ee
\be  \Delta s(p)^s_{{\rm sea}} =\cos^2 \phi \left[ -a \alpha^2
\right]+\sin^2 \phi \left[ -a \alpha^2 \right]\,. \label{dssea}
\ee
Similarly,  the contribution of the orbital angular momentum of
the ``quark sea'' to the magnetic moment of a given quark is
expressed as
\be
\mu(p)^s_{{\rm orbit}} = \frac{4}{3} [\mu (u_+ \rightarrow s_-)]-
\frac{1}{3} [\mu (d_+ \rightarrow s_-)]\,, \label{orbit p} \ee
where
\be
\mu(q_+ \rightarrow s_-) =\frac{e_{s}}{2M_q}\langle
l_q\rangle+\frac{e_{q}-e_{s}}{2M_{GB}}\langle l_{GB}\rangle\,. \ee
The quantities $(l_q, l_{GB})$ and $(M_q, M_{GB})$ are the orbital
angular momenta and masses of quark and GB, respectively. The
orbital angular momentum contribution to the magnetic moment due
to all the fluctuations is then given as \bea \mu(u_+ \rightarrow
s_-) &=& a \left[ -\frac{M^2_{\pi}}{2
{M}_{\pi}(M_u+{M}_{\pi})}-\frac{\alpha^2(M^2_{K}-3M^2_u)}{2
{M}_{K}(M_u+{M}_{K})}+\frac{(3+\beta^2+2 \zeta^2)M^2_{\eta}}{6
{M}_{\eta}(M_u+{M}_{\eta})} \right]{\mu}_N, \label{u} \\ \mu(d_+
\rightarrow s_-) &=& a \frac{M_u}{M_d} \left[ -\frac{\alpha^2
M^2_{K}}{2 {M}_{K}(M_d+{M}_{K})}-\frac{(\beta^2+2
\zeta^2)M^2_{\eta}}{12 {M}_{\eta}(M_d+{M}_{\eta})} \right]{\mu}_N,
\label{d}  \eea where  $M_{\pi}$, $M_{K}$ and $M_{\eta}$ are the
masses of pion, Kaon and $\eta$ respectively and $\mu_N$ is the
Bohr magneton. The strangeness contribution to the magnetic
moments of the neutron $n(ddu)$ as well as the decuplet baryons
$\Delta^{++}(uuu)$, $\Delta^{+}(uud)$, $\Delta^{o}(udd)$ and
$\Delta^{-}(ddd)$ can be calculated similarly.

The integral of the quark distribution functions \cite{xdep}
incorporating the strangeness content in the $\chi$CQM are
expressed as \cite{{cheng},hd} \be \bar u =\frac{1}{12}[(2
\zeta+\beta+1)^2 +20] a\,,~~~~  \bar d =\frac{1}{12}[(2 \zeta+
\beta -1)^2 +32] a\,,~~~~
 \bar s =\frac{1}{3}[(\zeta -\beta)^2 +9 {\alpha}^{2}] a\,, \label{barq}\ee
 \be u-\bar u=2\,,~~~~~d-\bar d=1\,,~~~~~s-\bar s=0\,.
\ee The integral of the quark distribution functions would
henceforth be referred to as `averaged' quark distribution
functions.

Similarly, the other important quantities having implications for
the strangeness contribution to the nucleon are the averaged quark
flavor fractions $f_q=\frac{q+\bar q}{\sum_{q} (q+\bar q)}$ which
are expressed in terms of the $\chi$CQM parameters as \bea f_u&=&
\frac{12+a(21+\beta^2+4 \zeta+4 \zeta^2+\beta(2+4
\zeta))}{3(6+a(9+\beta^2+6 \alpha^2+2 \zeta^2))}\,, \nonumber \\
f_d&=& \frac{6+a(33+\beta^2-4 \zeta+4 \zeta^2+\beta(-2+4
\zeta))}{3(6+a(9+\beta^2+6 \alpha^2+2 \zeta^2))}\,, \nonumber \\
 f_s&=& \frac{4a(\beta^2++9 \alpha^2-2
\beta\zeta+\zeta^2)}{3(6+a(9+\beta^2+6 \alpha^2+2 \zeta^2))}\,.
\label{fuds} \eea It is clear from the above expressions that the
non zero value of the parameters $a$, $\alpha$, $\beta$ and
$\zeta$ implies $f_s\neq 0$ as well as modify $f_u$ and $f_d$ due
to the strangeness contributions coming from the ``quark sea''.
Further, the  ratio of the functions
\be f_3= f_u-f_d\,, ~~~~f_8= f_u+f_d-2 f_s \,, \ee  and the
averaged ratios \be \frac{2 \bar s}{(u+d)}= \frac{4 a(9 \alpha^2
+\beta^2-2 \beta \zeta +\zeta^2)}{18+a(27+\beta^2+4 \beta\zeta+4
\zeta^2)}\,, ~~~ \frac{2 \bar s}{(\bar u+\bar d)}= \frac{4(9
\alpha^2 +\beta^2-2 \beta \zeta +\zeta^2)}{27+\beta^2+4
\beta\zeta+4 \zeta^2}\,, \ee have also been measured, therefore
providing an opportunity to check the strange quark content of the
nucleon.

The $\chi$CQM$_{{\rm config}}$ involves five parameters, four of
these $a$, $a \alpha^2$, $a \beta^2$, $a \zeta^2$ representing
respectively the probabilities of fluctuations to pions, $K$,
$\eta$, $\eta^{'}$, following the hierarchy $a > \alpha > \beta >
\zeta$, while the fifth representing the mixing angle. The mixing
angle $\phi$ is fixed from the consideration of neutron charge
radius \cite{yaouanc}, whereas for the other parameters we would
like to update our analysis using the latest data \cite{PDG}. In
this context, we find it convenient to use $\Delta u$, $\Delta_3$,
$\bar u-\bar d$ and $\bar u/\bar d$ as inputs with their latest
values given in Tables \ref{spin} and \ref{quark}. Before carrying
out the fit to the above mentioned parameters, we would like to
find their ranges by qualitative arguments. To this end, the range
of the symmetry breaking parameter $a$ can be easily found by
considering the spin polarization function $\Delta u$, by giving
the full variation to the parameters $\alpha$, $\beta$ and
$\zeta$, for example, one finds $0.10 \lesssim a \lesssim 0.14$.
The range of the parameter $\zeta$ can be found from averaged
$\bar u/\bar d$ using the latest experimental measurement
\cite{e866} and it comes out to be $-0.70 \lesssim \zeta \lesssim
-0.10$. Using the above found ranges of $a$ and $\zeta$ as well as
the latest measurement of $\bar u-\bar d$ asymmetry \cite{e866},
$\beta$ comes out to be in the range $0.2\lesssim \beta \lesssim
0.7$. Similarly, the range of $\alpha$ can be found by considering
the flavor non-singlet component $\Delta_3$ and it comes out to be
$0.2 \lesssim \alpha \lesssim 0.5$. After finding the ranges of
the symmetry breaking parameters, we have carried out a fine
grained analysis using the above ranges as well as considering
$\alpha\approx \beta$ by fitting $\Delta u$, $\Delta_3$ \cite{PDG}
as well as $\bar u-\bar d$, $\bar u/\bar d$ \cite{e866} leading to
$a=0.13$, $\zeta=-0.10$, $\alpha=\beta=0.45$ as the best fit
values. The parameters so obtained have been used to calculate the
spin polarization functions and the averaged quark distribution
functions. The calculated quantities pertaining to spin
polarization functions have been corrected by including the gluon
polarization effects \cite{cheng,ab} and symmetry breaking effects
\cite{cheng,johan}. Similarly, the averaged quark distribution
functions have been corrected by including the symmetry breaking
effects. The orbital angular momentum contributions to magnetic
moment  are characterized by the parameters of $\chi$CQM as well
as the masses of the GBs. For the $u$ and $d$ quarks, we have used
their most widely accepted values in hadron spectroscopy
\cite{cheng,yaoubook}, for example, $M_u=M_d=330$ MeV. For
evaluating the contribution of GBs, we have used its on mass shell
value in accordance with several other similar calculations
\cite{{mpi1}}.

In Tables \ref{spin}-\ref{quark}, we have presented the results of
our calculations pertaining to the strangeness dependent
parameters in $\chi$CQM$_{{\rm config}}$. For comparison sake, we
have also given the corresponding quantities in CQM. To begin
with, we first discuss the quality of fit pertaining to the spin
polarization functions. In Table \ref{spin}, we have presented the
strangeness incorporating spin polarization functions and the weak
axial vector couplings. Using $\Delta u$, $\Delta_3$ along with
$\bar u-\bar d$, $\bar u/\bar d$ from Table \ref{quark} as inputs,
we find that we are able to achieve a fairly good fit in the case
of spin polarization functions and the weak axial vector
couplings. In particular, the agreement in terms of the magnitude
as well as the sign in the case of $\Delta s$ is in good agreement
with the latest data \cite{smc,adams}.  The agreement in the case
of $\Delta_8$ and $\Delta_0$, which receives contribution from
$\Delta s$ also, not only justify the success of $\chi$CQM$_{{\rm
config}}$ but also strengthen our conclusion regarding $\Delta s$.
Similarly, the  agreement obtained in the case of the ratio $F/D$
again reinforces our conclusion that $\chi$CQM$_{{\rm config}}$ is
able to generate qualitatively as well as quantitatively the
requisite amount of strangeness in the nucleon.

In Table \ref{mag}, we have presented the spin and orbital
contributions pertaining to the strangeness magnetic moment of the
nucleon and $\Delta$ baryons. From the Table one finds that the
present result for the strangeness contribution to the magnetic
moment of proton looks to be in agreement with the most recent
results available for  $\mu(p)^s$
\cite{sample,mup-aw,global,lattice}. On closer examination of the
results, several interesting points emerge. The strangeness
contribution to the magnetic moment is coming from spin and
orbital angular momentum of the ``quark sea'' with opposite signs.
These contributions are fairly significant and they cancel in the
right direction to give the right magnitude to $\mu(p)^s$, For
example, the spin contribution in this case is $-0.09 \mu_N$ and
the contribution coming from the orbital angular momentum is $0.05
\mu_N$. These contributions cancel to give a small value for
$\mu(p)^s$ $-0.03 \mu_N$ which is consistent with the other
observed results. Interestingly, in the case of $\mu(n)^s$, the
magnetic moment is dominated by the orbital part as was observed
in the case of the total magnetic moments \cite{hdorbit} however,
the total strangeness magnetic moment is same as that of the
proton. Therefore, an experimental observation of this would not
only justify the Cheng-Li mechanism \cite{cheng} but would also
suggest that the chiral fluctuations is able to generate the
appropriate amount of strangeness in the nucleon. For the sake of
completeness, we have also presented the results of
$\mu(\Delta^{++})^s$, $\mu(\Delta^{+})^s$, $\mu(\Delta^{o})^s$,
$\mu(\Delta^{-})^s$ and here also we find that there is a
substantial contribution from spin and orbital angular momentum.
In general, one can find that whenever there is an excess of $d$
quarks the orbital part dominates, whereas when we have an excess
of $u$ quarks, the spin polarization dominates.

After finding that the  $\chi$CQM$_{{\rm config}}$ is able to give
a fairly good account of the spin dependent polarization
functions, in Table \ref{quark}, we have presented the results of
averaged quark distribution functions having implications for
strangeness in the nucleon. In line with the success of
$\chi$CQM$_{{\rm config}}$ in describing the spin dependent
polarization functions, in this case also we are able to give a
fairly good account of most of the measured values. The agreement
in the case of $\frac{2s}{u+d}$ and $\frac{f_3}{f_8}$ indicates
that, in the $\chi$CQM, we are able to generate the right amount
of strange quarks through chiral fluctuation. A refinement in the
case of the strangeness dependent quark ratio $\frac{2s}{\bar
u+\bar d}$ would have important implications for the basic tenets
of $\chi$CQM. The observed result for the case of $f_s$ in the
present case also indicates that the strange sea quarks play a
significant role in the nucleon. This is in agreement with the
observations of other authors \cite{cheng,gasser}.

To summarize, the  $\chi$CQM$_{{\rm config}}$ is able to provide a
fairly good description of the spin dependent polarization
functions as well as the averaged quark distribution functions
having implications for strangeness in the nucleon. It is able to
give a quantitative description of the important parameters such
as $\Delta s$, the weak axial vector couplings $\Delta_8$ and
$\Delta_0$, strangeness contribution to the magnetic moment
$\mu(p)^s$, the strange quark flavor fraction $f_s$, the
strangeness dependent ratios $\frac{2 \bar s}{u+d}$ and
$\frac{f_3}{f_8}$ etc.. In the case of $\mu(p)^s$, our result is
consistent with the latest experimental measurements as well as
with the other calculations. In conclusion, we would like to state
that at the leading order constituent quarks and the weakly
interacting Goldstone bosons constitute the appropriate degrees of
freedom in the nonperturbative regime of QCD and the ``quark sea''
generation in the $\chi$CQM$_{{\rm config}}$ through the chiral
fluctuation is the key in understanding the strangeness content of
the nucleon.

 \vskip .2cm
 {\bf ACKNOWLEDGMENTS}\\
H.D. would like to thank DST (Fast Track Scheme), Government of
India, for financial support and S. Singh, Head of Department, for
providing facilities to work in the department.

%\pagebreak

\begin{table}
\begin{center}
\begin{tabular}{cccc} \hline \hline
Parameter & Data  & CQM & $\chi$CQM$_{{\rm config}}$ \\   \hline

$\Delta u^*$ & 0.85 $\pm$ 0.05 \cite{smc} & 1.333   & 0.867
\\ $\Delta d$ & $-$0.41  $\pm$ 0.05 \cite{smc}   & $-$0.333 &
    $-$0.392 \\

$\Delta s$ & $-0.10 \pm 0.04 $ \cite{smc}   &   0 &
 $-0.08$
\\
& $-0.07 \pm 0.03 $ \cite{adams}   &        & \\

$\Delta_3^*$ & 1.267 $\pm$ 0.0035 \cite{PDG} & 1.666 &
 1.267\\ $\Delta_8$ & $0.58 \pm 0.025$ \cite{PDG} & 1
 & 0.59  \\ $\Delta_0$ & $0.19 \pm 0.025$ \cite{PDG} &
0.50     & 0.19
\\ $F/D$ & $0.575 \pm 0.016$ \cite{PDG} &
0.673     & 0.589
\\
\hline \hline {\small $*$ Input parameters} &&&

\end{tabular}
\end{center}
\caption{The calculated values of the strangeness dependent spin
polarization functions and weak axial vector couplings in the CQM
and $\chi$CQM$_{{\rm config}}$.}
 \label{spin}
\end{table}

\begin{table}
\begin{center}
\begin{tabular}{cccc} \hline \hline
Parameter & Data  & CQM & $\chi$CQM$_{{\rm config}}$ \\   \hline

$\mu(p)^s_{{\rm spin}}$, $\mu(p)^s_{{\rm orbit}}$ & $-$ &0, 0 &
$-0.09$, 0.05 \\ $\mu(p)^s$ & $ 0.37 \pm 0.26 \pm 0.20$
\cite{sample} & 0 & $-0.03$
\\  & $0.12 \pm 0.55 \pm 0.07$
\cite{global} & & \\

$\mu(n)^s_{{\rm spin}}$, $\mu(n)^s_{{\rm orbit}}$ & $-$ &0, 0
 & 0.06, $-0.09$ \\ $\mu(n)^s$ & $-$& 0 &  $-0.03$
\\

$\mu(\Delta^{++})^s_{{\rm spin}}$, $\mu(\Delta^{++})^s_{{\rm
orbit}}$ & $-$ & 0, 0  & $-0.29$, 0.18  \\ $\mu(\Delta^{++})^s$ &
$-$ & 0 &
 $-0.11$ \\

$\mu(\Delta^{+})^s_{{\rm spin}}$, $\mu(\Delta^{+})^s_{{\rm
orbit}}$ & $-$ & 0, 0 & $-0.14$,  0.11 \\ $\mu(\Delta^{+})^s$ &
$-$ & 0 & $-0.03$ \\

$\mu(\Delta^{o})^s_{{\rm spin}}$, $\mu(\Delta^{o})^s_{{\rm
orbit}}$ & $-$ & 0, 0  &   $-0.04$, $-0.03$ \\ $\mu(\Delta^{o})^s$
& $-$ & 0 & $-0.07$ \\

$\mu(\Delta^{-})^s_{{\rm spin}}$, $\mu(\Delta^{-})^s_{{\rm
orbit}}$ & $-$ & 0, 0 & $-0.09$, 0.15 \\ $\mu(\Delta^{-})^s$ & $-$
& 0 & $0.06$
\\  \hline \hline

\end{tabular}
\end{center}
\caption{The calculated values of the strangeness contribution to
the magnetic moment of nucleon and $\Delta$ decuplet baryons in
the CQM and $\chi$CQM$_{{\rm config}}$.}
 \label{mag}
\end{table}

\begin{table}
\begin{center}
\begin{tabular}{cccc} \hline \hline
Parameter & Data  & CQM & $\chi$CQM$_{{\rm config}}$ \\   \hline

$\bar s$ & $-$   &   0 & 0.10 \\

$\bar u-\bar d^*$ & $-0.118 \pm$ 0.015 \cite{e866} & 0   &
 $-0.118$
\\

$\bar u/\bar d^*$ & 0.67 $\pm$ 0.06 \cite{e866}  & 0
 & 0.66\\

$\frac{2 \bar s}{u+d}$ & 0.099$^{+0.009}_{0.006}$ \cite{nutev} & 0
&
 0.09
\\

$\frac{2 \bar s}{\bar u+\bar d}$ & 0.477$^{+0.063}_{0.053}$
\cite{nutev} & 0  &  0.44
\\

$f_s$ &  0.10 $\pm$ 0.06 \cite{nutev}  &  $-$ & $0.08$
\\
$f_3$  &$-$ & $-$  & 0.21\\

$f_8$  &$-$ & $-$ & 1.03\\ $f_3/f_8$ & 0.21 $\pm$ 0.05
\cite{nutev} & 0.33 &   0.20 \\ \hline \hline {\small $*$ Input
parameters} &&&

\end{tabular}
\end{center}
 \caption{The calculated values of the strangeness dependent averaged quark flavor
distribution functions and related parameters in the CQM and
$\chi$CQM$_{{\rm config}}$.} \label{quark}
\end{table}

\end{document}